\documentclass[english]{article}
\usepackage[T1]{fontenc}
\usepackage[latin9]{inputenc}
\usepackage{xcolor}
\usepackage{amstext}
\usepackage[all]{xy}
\PassOptionsToPackage{normalem}{ulem}
\usepackage{ulem}

\makeatletter

\providecolor{lyxadded}{rgb}{0,0,1}
\providecolor{lyxdeleted}{rgb}{1,0,0}

\DeclareRobustCommand{\lyxsout}[1]{\ifx\\#1\else\sout{#1}\fi}

\date{}
\usepackage{cite}
\usepackage{geometry}

\makeatother

\usepackage{babel}
\begin{document}
\title{On Joint Distributions, Counterfactual Values, and Hidden Variables
in Understanding Contextuality}
\author{Ehtibar Dzhafarov\\Purdue University}
\maketitle
\begin{abstract}
This paper deals with three traditional ways of defining contextuality:
(C1) in terms of (non)existence of certain joint distributions involving
measurements made in several mutually exclusive contexts; (C2) in
terms of relationship between factual measurements in a given context
and counterfactual measurements that could be made if one used other
contexts; and (C3) in terms of (non)existence of ``hidden variables''
that determine the outcomes of all factually performed measurements.
It is generally believed that the three meanings are equivalent, but
the issues involved are not entirely transparent. Thus, arguments
have been offered that C2 may have nothing to do with C1, and the
traditional formulation of C1 itself encounters difficulties when
measurement outcomes in a contextual system are treated as random
variables. I show that if C1 is formulated within the framework of
the Contextuality-by-Default (CbD) theory, the notion of a probabilistic
coupling, the core mathematical tool of CbD, subsumes both counterfactual
values and ``hidden variables''. In the latter case, a coupling
itself can be viewed as a maximally parsimonious choice of a hidden
variable. 
\end{abstract}

\section{Introduction}

The aim of this paper is to consider three historically established
ways of understanding (non)contextuality, and relate them to each
other from the vantage point of the Contextuality-by-Default (CbD)
theory. The reader unfamiliar with CbD can find its latest version
in Ref. \cite{DzhCerKuj2017} (and additional details, arguments,
and proofs in Refs. \cite{DzhKuj2014,Dzh2017Nothing,DzhKuj2017Fortsch,KujDzhLar2015,KujDzhProof2016}).
My main point is that the three approaches in question can be viewed
as variants or interpretations of the core mathematical tool of CbD
--- probabilistic couplings.

The first of the three meanings of contextuality considered in this
paper can be called \emph{joint-distributional}. It was introduced
by Suppes and Zanotti \cite{SuppesZanotti1981} and Fine \cite{Fine1982},
within the conceptual framework of establishing its equivalence to
the more traditional at the time ``hidden-variable'' meaning of
contextuality, as discussed below. Contexts are defined as conditions,
or arrangements under which one performs one's measurements (including
but not reduced to what other measurements are performed together
with a given one). A system of measurements made in varying, mutually
exclusive contexts is noncontextual if the random variables representing
all these measurements can be considered jointly distributed. This
seems to have become a common way of understanding contextuality,
as evidenced by numerous contemporary works \cite{AbramBarbMans2011,AbramskyBrand2011,Araujoetal2013,BudroniBook2016,Cabello2013,Klyachko2008,Kurzynski2014,Kurzynskietal2012,LiangSpekkensWiseman2011}.
The CbD theory belongs to the same category, and its specific feature
is that all measurement outcomes are consistently treated as contextually
labelled random variables. The use of contextual labeling means that
measurements made in different contexts are always represented by
different random variables, even if they measure the same property.
As a result, the possibility of imposing a joint distribution on all
random variables in a system is not a restrictive requirement. In
CbD, therefore, one is interested in the existence of not just some
but only specific joint distributions, those satisfying certain requirements.
To match the traditional understanding, derived from Fine's and Suppes
and Zanotti's work, the requirement should be that, in the joint distribution
imposed on the system, the variables measuring the same property in
different contexts should always have equal values (which is achievable
only if the random variables in play satisfy the no-disturbance constraint,
as discussed below).

The second meaning of contextuality can be called \emph{counterfactual}.
It is formulated in terms of whether an outcome of a factual measurement
made in some context would have been the same had it been made in
another context. It seems that most contemporary researchers take
this counterfactual formulation as being equivalent to the joint-distributional
meaning of contextuality mentioned above (and to the ``hidden-variable''
meaning, mentioned below). This is apparent, e.g., in Liang, Speckens,
and Wisemen's comprehensive introduction to contextuality \cite{LiangSpekkensWiseman2011}.
However, this view is not universally accepted. Thus, Griffiths \cite{Griffiths2017}
calls the counterfactual meaning of contextuality \emph{Bell-contextuality},
and argues that any system is Bell-noncontextual (see Griffiths's
paper in the present issue, \cite{Griffiths2019}). This means that,
in Griffiths's opinion, it is always true that had one measured a
factually measured property in another context, the result would have
been the same; and that this has nothing to do with the existence
or nonexistence of joint distributions in the first meaning of contextuality.

The third way of defining (non)contextuality was historically the
first. Contextuality (without using this term) was introduced in quantum
physics through the notion of hidden variables, primarily by Bell
\cite{Bell1964,Bell1966} and Kochen and Specker \cite{KochenSpecker1967}.
In particular, Bell demonstrated that one could meaningfully address,
using only observable measurements, the question famously discussed
in Bohr's \cite{Bohr1935} critique of Einstein, Podolsky, and Rosen
\cite{EPR1935}. This question is whether all measurement outcomes
in a system of measurements can be presented as being determined by
some ``hidden'' random variable in a context-independent way, i.e.,
using context-independent mappings of the values of this hidden variable
into the values of the observed measurement outcomes. The question
has beens historically formulated in terms of ``realism'', the existence
of hidden variables of which all observable outcomes of measurements
are functions, and ``(non)locality'', the (in)dependence of these
functions on the contexts. (To include systems in which spatial separation
plays no role, e.g., of the Kochen-Specker variety, the term ``locality''
should be replaced with the broader term ``context-independence''.)
Quantum mechanics is usually said to exclude the conjunction of realism
and context-independence, but the culprit in this conjunction is not
agreed on by all. Thus, Leggett \cite{Leggett20031} and Gröblacher,
Paterek, Kaltenbaek, Brukner, \.{Z}ukowski, Aspelmeyer, and Zeilinger
\cite{Zeilinger2007} show that quantum mechanics rules out realism
in conjunction with certain forms of context-dependent mapping.

In this paper, I uphold the prevalent view that all three meanings
of contextuality are equivalent. We will see that counterfactual definiteness
and the existence of hidden variables can be viewed as philosophically
and/or physically laden ways of speaking of probabilistic couplings,
the notion that lies at the heart of CbD.

CbD is usually taken to be useful for \emph{inconsistently connected
systems} (systems with ``disturbance''), where measurements of the
same property in different contexts may have differently distributed
outcomes \cite{ariasetal2015,crespietal2017,fluhmanetal2018,malinowskietal2018,zhanetall2017,KujDzhLar2015}.
However, if measurements are treated as random variables within the
framework of classical probability theory, CbD offers considerable
conceptual clarity even for consistently connected systems, those
with no ``disturbance''. To illustrate this, I focus on such systems
throughout most of this paper. 

The following three aspects of our discussion should be kept in mind.
First, the term \emph{measurement} can be replaced with any procedure
with generally random outcomes, e.g., responses of a biological organism
to stimuli. For this reason I prefer in the following to use the standard
CbD term ``content'' (of a random variable) in place of the ``property
being measured''. Second, the analysis is entirely within the framework
of classical probability theory, with classical understanding of random
variables. In particular, random variables may but need not be related
to observables in Hilbert space. Third, the arguments I present in
favor of using CbD should not be misconstrued as criticism of other
contemporary approaches to (non)contextuality, such as presented in
Refs. \cite{AbramBarbMans2011,Araujoetal2013,BudroniBook2016,Khrennikov2009,LiangSpekkensWiseman2011}.
In particular, the terms ``traditional'' and ``historical'' used
in describing positions contrasted with CbD refer primarily to the
literature of the last century. 

\section{\label{sec:Preliminaries:-Terminology}Preliminaries: Terminology
and notation}

Consider an experiment consisting in measuring several properties,
generically called \emph{contents}, under various conditions, called
\emph{contexts}. The contents form a set $Q$, the contexts form a
set $C$, and in each context $c\in C$ one jointly measures some
subset $Q_{c}$ of the properties $Q$, with $Q_{c}\cap Q_{c'}$ generally
nonempty. If $q\in Q_{c}$, the result of measuring this content $q$
in a context $c$ is a random variable $R_{q}^{c}$. The set of the
random variables double-labeled in this way is a \emph{system} (of
random variables).

The random variables belonging to the same context form a set of jointly
distributed random variables 
\begin{equation}
\left\{ R_{q}^{c}:q\in Q_{c}\right\} .\label{eq: bunch}
\end{equation}
Conceptually, the joint distribution of context-sharing variables
means that they can be presented as measurable functions on one and
the same \emph{domain probability space} (sometimes also referred
to as \emph{sample space}). Equivalently, and more conveniently for
our purposes, this means that one can choose a random variable $H_{c}$
for each context $c$, and functions $f_{q}^{c}$ , such that 
\begin{equation}
f_{q}^{c}\left(H_{c}\right)=R_{q}^{c},\;q\in Q_{c},c\in C.\label{eq: trivial}
\end{equation}
If the sample space for the context-sharing variables in (\ref{eq: bunch})
is specified, $H_{c}$ can be chosen as the identity function on the
same space, in which case $f_{q}^{c}$ is the measurable function
defining $R_{q}^{c}$. However, a practical, and most economic, choice
of $H_{c}$ is (\ref{eq: bunch}) itself, which is a random variable
in its own right.\footnote{As mentioned in Section 5, the distinction between a ``single''
variable and a set of jointly distributed variables is purely representational
and dispensable.} With this choice, $f_{q}^{c}$ is the componentwise projection function.

By contrast with (\ref{eq: bunch}), any two random variables picked
from different contexts are \emph{stochastically unrelated}, even
if sharing a content: $H_{c}$ and $H_{c'}$ for $c\not=c'$ have
no joint distribution (are defined on distinct sample spaces).

As mentioned above, we focus on \emph{consistently connected} \emph{systems},
defined by the following property: if $q\in Q_{c}\cap Q_{c'}$ ($c,c'\in C$),
then the distribution of $R_{q}^{c}$ is the same as the distribution
of $R_{q}^{c'}$. This is written as 
\begin{equation}
R_{q}^{c}\sim R_{q}^{c'}.
\end{equation}
That is, a random variable's distribution in a consistently connected
system is determined by the variable's content only.\footnote{This property is known in physics under a variety of names, such as
no-signaling, no-disturbance, parameter invariance, etc. \cite{Cereceda2000}.
In psychology it is known under the name of simple marginal selectivity.
The term ``consistent connectedness'' is less connotationally loaded
than most of these terms. It is a naturally formed term in CbD, because
contexts in this theory are represented by isolated ``islands''
of random variables, and they are only ``connected'' to each other
by the fact that some variables in different contexts have the same
content. The set of all variables $R_{q}^{c}$ with the same content
$q$ is then natural to call a \emph{connection} \cite{DzhKuj2016}.}

As an example I will use the following system:
\begin{equation}
\begin{array}{|c|c|c||c}
\hline R_{1}^{1} & R_{2}^{1} &  & c=1\\
\hline  & R_{2}^{2} & R_{3}^{2} & c=2\\
\hline R_{1}^{3} &  & R_{3}^{3} & c=3\\
\hline\hline q=1 & q=2 & q=3
\end{array}.\label{eq: matrix 3}
\end{equation}
Consistent connectedness means here that $R_{1}^{1}\sim R_{1}^{3}$,
$R_{2}^{1}\sim R_{2}^{2}$, and $R_{3}^{2}\sim R_{3}^{3}$. Because
the two variables within each context are measured together, they
are \emph{operationally }(or\emph{ empirically}) jointly distributed.\footnote{The term ``measured together'' usually means simultaneously, but
it could be any empirical scheme by which the values of two variables
are paired. They can be measured at different times, as in the Leggett-Garg-type
experiments \cite{LeggettGarg1985,BudroniBook2016}, and they can
be related to the same or different physical objects. This is one
reason my analysis is not critically related to such issues as commutativity
of the quantum operators generating the random variables.} By contrast, random variables picked from different contexts, say
$R_{1}^{1}$ and $R_{3}^{3}$ or $R_{1}^{1}$ and $R_{1}^{3}$, are
\emph{stochastically unrelated}: there is no empirical procedure for
pairing their values; they can be paired ``on paper'', but not uniquely,
with no particular way being privileged.

I will use the example of system (\ref{eq: matrix 3}) in the remainder,
sometimes mentioning and sometimes only implying a generalization
to any system of random variables. However, the generalization is
trivial only if we confine our discussion to systems of random variables
with finite numbers of contents and contexts, and to random variables
that are \emph{categorical}, i.e., have finite numbers of values.
The latter is not a restriction for the CbD approach, in which all
random variables should be replaced with sets of jointly distributed
\emph{dichotomous} random variables before contextuality analysis
can be applied \cite{DzhCerKuj2017} (we need not, however, discuss
this construction in this paper). 

\section{Joint-distributional understanding: Identically connected couplings }

The first meaning of contextuality is based on the obvious fact that
one can consider a multitude of ways the six random variables in (\ref{eq: matrix 3})
could be jointly distributed ``on paper'' (knowing that they are
not jointly distributed de facto). This formulation (with the words
``could be'') is hinting at counterfactuality, but we need not go
that way: mathematically, we simply consider all sextuples of jointly
distributed random variables
\begin{equation}
\begin{array}{|c|c|c|}
\hline S_{1}^{1} & S_{2}^{1} & \\
\hline  & S_{2}^{2} & S_{3}^{2}\\
\hline S_{1}^{3} &  & S_{3}^{3}
\\\hline \end{array},\label{eq: matrix 3 coupling}
\end{equation}
with the same row-wise distributions as in (\ref{eq: matrix 3}).
Any such a sextuple is a \emph{probabilistic coupling} of the system
(\ref{eq: matrix 3}), and the set of possible couplings is always
nonempty. More generally, given a set $\mathcal{X}$ of random variables,
its coupling is defined as a set $X$ of jointly distributed random
variables, in a bijective correspondence with $\mathcal{X}$, such
that for any $\mathcal{Y}\subseteq\mathcal{X}$, if the elements of
$\mathcal{Y}$ are jointly distributed, then the corresponding subset
$Y\subseteq X$ has the same distribution. In particular, corresponding
elements of $\mathcal{X}$ and $X$ are identically distributed.\footnote{This definition is modified with respect to the standard one \cite{Thorisson}
to better suit contextuality analysis.}

In contextuality analysis of a consistently connected system one is
interested in whether among all possible couplings of a system of
random variables one can find one with a special property. This property
is that in each connection, i.e. each column of (\ref{eq: matrix 3 coupling}),
the two random variables are equal to each other with probability
1. If such a coupling exists (in CbD it is called\emph{ identically
connected}), then the system (\ref{eq: matrix 3}) is considered \emph{noncontextual}.
Otherwise it is \emph{contextual}. Thus, if the variables in the system
(\ref{eq: matrix 3}) are dichotomous, $+1/-1$, then it is known
\cite{KujDzhProof2016,Araujoetal2013} that this system is contextual
if and only if 
\begin{equation}
\max_{\textnormal{odd \# of }-\textnormal{'s}}\left(\pm\left\langle R_{1}^{1}R_{2}^{1}\right\rangle \pm\left\langle R_{2}^{2}R_{3}^{2}\right\rangle \pm\left\langle R_{3}^{3}R_{1}^{3}\right\rangle \right)>1,\label{eq: suppes-zanotti}
\end{equation}
where the maximum is taken over all combinations with odd numbers
of minus signs (1 or 3). Examples of such systems are readily constructed. 

Traditionally, the system (\ref{eq: matrix 3}) would be presented
as
\begin{equation}
\begin{array}{|c|c|c||c}
\hline R_{1} & R_{2} &  & c=1\\
\hline  & R_{2} & R_{3} & c=2\\
\hline R_{1} &  & R_{3} & c=3\\
\hline\hline q=1 & q=2 & q=3
\end{array},\label{eq: matrix 3 bad}
\end{equation}
with overlapping sets of random variables (here, each random variable
occurs in two different contexts). This is the case, e.g., in what
seems to be historically very first joint-distributional analysis
of contextuality (without using this term), the 1981 paper by Suppes
and Zanotti \cite{SuppesZanotti1981}. The central theorem in that
paper says (mutatis mutandis):\footnote{Compared to the original formulation, notation is changed, an unnecessary
constraint is removed, and the inequality is replaced with an equivalent
one to make it comparable to (\ref{eq: suppes-zanotti}).} 
\begin{quote}
Suppes-Zanotti's Theorem. Let $R_{1},R_{2},R_{3}$ be random variables
with possible values 1 and -1. Then a necessary and sufficient condition
for the existence of a joint probability distribution of the three
random variables is
\begin{equation}
\max_{\textnormal{odd \# of }-\textnormal{'s}}\left(\pm\left\langle R_{1}R_{2}\right\rangle \pm\left\langle R_{2}R_{3}\right\rangle \pm\left\langle R_{3}R_{1}\right\rangle \right)\leq1.\label{eq: suppes-zanotti bad <}
\end{equation}
\end{quote}
While the necessity part of this statement is straightforward, the
sufficiency part encounters difficulties. The reason for this is that
the relation of being jointly distributed is ``agglutinative'',
in the following sense:
\begin{description}
\item [{(Agglutinativity)}] given sets $A$ and $B$ of jointly distributed
random variables, if $A\cap B\not=\emptyset$, then $A\cup B$ is
a set of jointly distributed random variables.\footnote{In a previous publication \cite{DzhKuj2017Fortsch} this property
was erroneously called ``transitivity''. When applied to three random
variables, $X,Y,Z$, transitivity means that joint distributions of
$\left(X,Y\right)$ and $\left(Y,Z\right)$ implies that of $\left(X,Z\right)$.
Agglutinativity means that joint distribution of $\left(X,Y\right)$
and $\left(Y,Z\right)$ implies that of $\left(X,Y,Z\right)$. This
implies transitivity but is not equivalent to it.}
\end{description}
This property holds essentially by definition of a random variable.
It follows that for $R_{1},R_{2},R_{3}$ to be jointly distributed
it is sufficient that at least two of the expected values $\left\langle R_{1}R_{2}\right\rangle ,\left\langle R_{2}R_{3}\right\rangle ,\left\langle R_{3}R_{1}\right\rangle $
be well-defined, i.e. at least two of the pairs $\left(R_{1},R_{2}\right),\left(R_{2},R_{3}\right),\left(R_{3},R_{1}\right)$
be jointly distributed. But the latter would be the case even if 
\begin{equation}
\max_{\textnormal{odd \# of }-\textnormal{'s}}\left(\pm\left\langle R_{1}R_{2}\right\rangle \pm\left\langle R_{2}R_{3}\right\rangle \pm\left\langle R_{3}R_{1}\right\rangle \right)>1,\label{eq: suppes-zanotti bad >}
\end{equation}
in which case $R_{1},R_{2},R_{3}$ cannot be jointly distributed.
This contradiction shows that a correct formulation of Suppes-Zanotti's
theorem should have been as follows:
\begin{quote}
Let $R_{1},R_{2},R_{3}$ be random variables with possible values
1 and -1. Then a necessary and sufficient condition for the existence
of a joint probability distribution of the three random variables
is the existence of a joint distribution of any two of the three pairs
$\left(R_{1},R_{2}\right),\left(R_{2},R_{3}\right),\left(R_{3},R_{1}\right)$.
If this is the case, (\ref{eq: suppes-zanotti bad <}) is satisfied.
\end{quote}
However, then it follows that (\ref{eq: suppes-zanotti bad >}) cannot
ever hold. Put differently, if (\ref{eq: matrix 3 bad}) is a \emph{system}
(implying, in particular, that the joint distributions within contexts
are well-defined), then this system can only be noncontextual. Therefore,
if it happens that (\ref{eq: suppes-zanotti bad >}) holds for this
system, then one has a true contradiction on one's hands, and this
contradiction cannot be resolved within the framework of (\ref{eq: matrix 3 bad}).
It can only be resolved by explicating and rejecting some hidden assumptions
-- and in this case the culprit is the assumption that the random
variables measuring the same content in different contexts are the
same. 

This problem is ubiquitous in the traditional literature succeeding
Ref. \cite{SuppesZanotti1981}, although its critical analysis is
complicated by the fact that many authors would refer to elements
of (\ref{eq: matrix 3 bad}) as measurements or observables rather
than random variables. I take it as a given, however, that the notion
of a distribution of $R_{i}$, or the probability of $R_{i}$ being
equal to some value, can only be used if $R_{i}$ is a random variable.
With this in mind, the contradiction just described can only be resolved
by using contextual notation, and CbD offers a straightforward way
of doing this. However, contextual notation can also be applied to
probabilities rather than random variables per se, and this seems
to be the way chosen in some of the contemporary literature. Thus,
Khrennikov \cite{Khrennikov2005,Khrennikov2009} proposes labeling
of the form $\Pr\left[R_{i}=r\,|\,c=j\right]$, calling this ``contextual
probabilities'' and warning against identifying them with conditional
probabilities. Liang, Speckens, and Wisemen \cite{LiangSpekkensWiseman2011}
use essentially the same notation (if one considers quantum preparations
part of contexts). Abramsky and colleagues \cite{AbramBarbMans2011,AbramskyBrand2011}
developed a similar system in which $e_{c}$ denotes the joint probability
of all random variables in context $c$, and distributions of their
subsets are treated as specializations: e.g., the distribution of
$R_{1}$ in $c=3$ would be denoted $e_{c=3}|_{q=1}$. Contextual
notation for probabilities, the same as CbD's contextual notation
for random variables, allows one to avoid the difficulties related
to the agglutinativity. 

\section{Counterfactual approach}

The second meaning of (non)contextuality is predicated on an affirmative
answer to the following question:
\begin{description}
\item [{(Q1:$\mathbf{\:\textbf{counterfactual definiteness}}$)}] when
one makes measurements in a given context, can one meaningfully speak
of what the outcomes of measurements would have been had one chosen
another context?
\end{description}
Using our example system (\ref{eq: matrix 3}), if the chosen context
is $c=2$, one records the values of $R_{2}^{2}$ and $R_{3}^{2}$.
Is it meaningful to ask what the recorded values would have been had
we chosen $c=1$ or $c=3$ instead of $c=2$? If the answer is negative,
there is nothing more to discuss, and the counterfactual meaning of
contextuality cannot be formulated. A positive answer means that whenever
a measurement is being made,\emph{ all random variables in all contexts
can be thought of as having definite values}. It makes no difference
for any possible consequences whether this assignment of values is
understood epistemologically (the experimenter can always assign these
counterfactual values, perhaps not uniquely) or ontologically (the
random variables in counterfactual contexts have true values, but
being unknown in principle, one should consider possibilities, perhaps
more than one, of what these ``true'' values could be). 

Assuming the answer to Q1 is positive, one can ask the next, critical
question:
\begin{description}
\item [{(Q2:$\mathbf{\:\textbf{counterfactual identity}}$)}] is it possible
that in this assignment of values to all random variables in the system
any two content-sharing random variables $R_{q}^{c}$ and $R_{q}^{c'}$
be always assigned the same value?
\end{description}
If this question, too, is answered in the affirmative, i.e., if there
is an assignment of values that only depends on the variables' contents,
rather than also on their contexts, the system is deemed noncontextual
in the ``counterfactual sense''.

Note: in our understanding of Q2, the variables $R_{q}^{c}$ and $R_{q}^{c'}$
measure the same content in two distinct contexts, one of which may
be, but need not be factual. One might object to this and argue that
one should only be interested in this question if one of the contexts
$c,c'$ is the factual context. A comparison of two counterfactual
assignments, one might insist, is of no interest. I can see no convincing
justification for imposing this restriction. Using our example (\ref{eq: matrix 3}),
if one meaningfully contemplates a counterfactual value of $R_{2}^{1}$
when the measurements are factually made in context $c=2$, one should
also be able to consider counterfactual value of $R_{1}^{1}$: after
all, $R_{2}^{1}$ and $R_{1}^{1}$ are jointly distributed, it may
even be the case that one of them is uniquely determined by the other,
e.g., $R_{2}^{1}=-R_{1}^{1}$. But then, if one can speak of the value
that $R_{1}^{1}$ would have had if one measured $q=1$ in context
$c=1$ instead of measuring $q=2$ in context $c=2$, then one can
also speak of what the value of $R_{1}^{3}$ would have been if the
same $q=1$ were measured in context $c=3$. And the requirement that
the assignment of values should only depend on content would then
dictate that $R_{1}^{3}$ be assigned the same value as $R_{1}^{1}$.

The picture we arrive at now is: irrespective of what factual measurements
are made, all random variables are assigned values, and we ask whether
this can be done so that any two $R_{q}^{c}$ and $R_{q}^{c'}$ be
assigned the same value. Assume that this is possible, i.e., the system
is noncontextual in the counterfactual sense. Then $R_{q}^{c}$ and
$R_{q}^{c'}$, being always equal to each other, are jointly distributed.
By the agglutinativity of the relation of being jointly distributed,
this means that all random variables in the system are jointly distributed.
This is obvious in our example, where the identity requirement across
contexts together with the joint distributions within contexts yields
the following graph of pairwise jointly distributed random variables: 

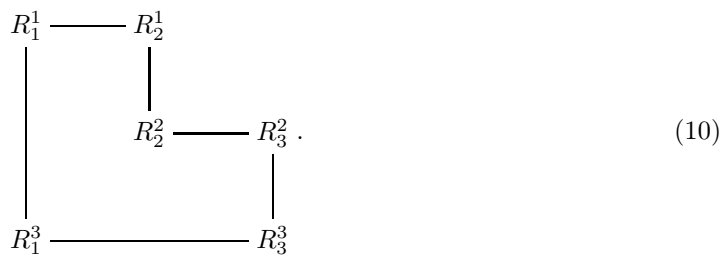
\begin{equation}
\vcenter{\xymatrix@C=1cm{R_{1}^{1}\ar@{-}[r] & R_{2}^{1}\ar@{-}[d]\\
 & R_{2}^{2}\ar@{-}[r] & R_{3}^{2}\ar@{-}[d]\\
R_{1}^{3}\ar@{-}[uu] &  & R_{3}^{3}\ar@{-}[ll]
}
}.\label{eq: diagram}
\end{equation}
Any path in this graph that includes all nodes suffices to establish
that the random variables are jointly distributed. Even easier, consider
contexts as nodes of a graph and connect two contexts by an edge if
they involve at least one common content. In our example it would
look like this:
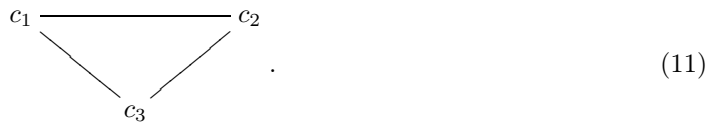
\begin{equation}
\vcenter{\xymatrix@C=1cm{c_{1}\ar@{-}[rr] &  & c_{2}\ar@{-}[dl]\\
 & c_{3}\ar@{-}[ul]
}
}.\label{eq: diagram-1}
\end{equation}
If such a graph is constructed for an arbitrary system, and if, as
in our example, it is \emph{connected}, then in this system all random
variables are jointly distributed. Systems whose context graphs are
not connected do not pose difficulties as they can (and should) always
be studied as several unrelated to each other systems with connected
graphs.

It is now obvious that a system that is noncontextual (or contextual)
in the the counterfactual sense is precisely a system for which there
exists (respectively, does not exist) an identically connected coupling.
The two meanings coincide. One can even say that the notion of a coupling
is nothing but a rigorous mathematical meaning of the ``counterfactual
sense''. Instead of making the assumption, some would say distinctly
metaphysical in flavor, that a random variable has a definite value
even if unmeasured, one can state as a fact, with no assumptions involved,
that one can impose a joint distribution on (construct a coupling
of) all random variables in the system. And then one can ask whether
among all such couplings one can find an identically connected one.
Counterfactual statements can be rigorously formalized, but most would
agree that their logical status is more involved than that of factual
statements. ``The system has a coupling with properties $X$'' is
more ``ordinary'' a statement than ``Had we measured $A$ in another
context its value would have been $x$.'' As stated in a Stanford
Encyclopedia of Philosophy article \cite{William}, ``Philosophers,
linguists, and psychologists remain fiercely divided on how to best
understand counterfactuals.'' It is hard to be divided over the notion
of a coupling. 

\section{Hidden variables with context-independent mapping}

As mentioned in Section \ref{sec:Preliminaries:-Terminology}, several
random variables are jointly distributed if and only if they are functions
of one and the same random variable. Thus, because in each content
$c$ all random variables $R_{q}^{c}$ ($q\in Q_{c}$) are jointly
distributed in the operational sense (measured ``together''), one
can define a random variable $H_{c}$ and functions $f_{q}^{c}$ ($q\in Q_{c}$)
such that (\ref{eq: trivial}) holds. This is a context-specific hidden-variable
construction, and it is obviously nonrestrictive, applicable to any
system. The third meaning of (non)contextuality we are discussing
now is about the possibility of a single hidden variable for all contexts,
and context-independent functions that map it into the random variables
comprising the system. In other words, one asks whether there is a
random variable $H$ and a set of functions $f_{q}$ ($q\in Q$) such
that for any $c\in C$ and $q\in Q_{c}$, 
\begin{equation}
R_{q}^{c}=f_{q}\left(H\right).\label{eq: HV hypothesis}
\end{equation}
If (and only if) such a construction is possible, the system is noncontextual
in the ``hidden variable'' sense.

Applied to our example, the question is about the possibility of replacing
(\ref{eq: matrix 3}) with
\begin{equation}
\begin{array}{|c|c|c||c}
\hline f_{1}\left(H\right) & f_{2}\left(H\right) &  & c=1\\
\hline  & f_{2}\left(\text{\ensuremath{H}}\right) & f_{3}\left(\text{\ensuremath{H}}\right) & c=2\\
\hline f_{1}\left(\text{\ensuremath{H}}\right) &  & f_{3}\left(\text{\ensuremath{H}}\right) & c=3\\
\hline\hline q=1 & q=2 & q=3
\end{array}.\label{eq: HV matrix}
\end{equation}
At every measurement, the variable $H$ has some value, and all six
random variables, irrespective of what context $c$ is being factually
measured, are determined by this value. At that, the values of any
two random variables measuring the same content are always equal (because
the functions are labeled by their contents only). It is easy to see
that this is precisely the same as the existence, for any factual
measurement, of the assignment of values to all random variables in
the system, such that $R_{q}^{c}=R_{q}^{c'}$ for any $q\in Q_{c}\cap Q_{c'}$.
Of course, in CbD, (\ref{eq: HV hypothesis}) has to be replaced with
\begin{equation}
R_{q}^{c}\sim f_{q}\left(H\right)=S_{q},\label{eq: HV correct}
\end{equation}
i.e. (\ref{eq: HV matrix}) represents a coupling of the system rather
than the system itself. 

This completes the demonstration that the three meanings of (non)contextuality
considered in this paper are subsumed by the notion of a coupling
in the CbD sense. 

There is also a ``stochastic'' version of the hidden-variable hypothesis,
in which each variable $R_{q}^{c}$ is a function $f_{q}$ of a common
source of randomness $H$ and some specific source of randomness $V_{q}$
(context-independent),
\begin{equation}
R_{q}^{c}\sim f_{q}\left(H,V_{q}\right).
\end{equation}
This version, however, is immediately reduced to the previous, ``deterministic''
version (\ref{eq: HV hypothesis}), on renaming 
\begin{equation}
\left(H,\left\{ V_{q}\right\} _{q\in Q}\right)\mapsto H'.
\end{equation}

The existence of the single underlying random variables $H$ is sometimes
referred to as ``realism'', whereas the context-independence of
the mappings $f_{q}$ is the generalization of what is traditionally
referred to as ``locality'' (when applied to spatially distributed
systems of particles). It is worth noting that of these two requirements
for a hidden variable theory, it is only context-independence of mappings
that has a restrictive effect. Indeed, the variables $H_{c}$ in (\ref{eq: trivial}),
of which we know that it is universally applicable, can always be
coupled in a variety of ways, e.g. independently. In our example,
let us replace $H_{c=1},H_{c=2},H_{c=3}$ with a triple of jointly
distributed, e.g. independent, random variables $\left(G_{c=1},G_{c=2},G_{c=3}\right)$
such that $G_{c}\sim H_{c}$. Being jointly distributed, $G_{c=1},G_{c=2},G_{c=3}$
can be presented as functions of some random variables $G$, 
\[
G_{c}=g_{c}\left(G\right),\;c=1,2,3.
\]
This variable $G$ can be chosen, e.g., as the vector 
\begin{equation}
G=\left(G_{c=1},G_{c=2},G_{c=3}\right),\label{eq: U3}
\end{equation}
in which case
\begin{equation}
R_{q}^{c}\sim f_{q}^{c}\left(g_{c}\left(G\right)\right)=h_{q}^{c}\left(G\right),\;c=1,2,3,q=1,2,3,\label{eq: with c}
\end{equation}
where $g_{c}$ is the $c$th projection function. Denoting
\begin{equation}
h_{q}^{c}\left(G\right)=S_{q}^{c},
\end{equation}
we form a coupling of the system. In accordance with \cite{Leggett20031},
this ``realist'' construction is completely nonrestrictive, applicable
to any system. This does not exclude the possibility that some special
cases of context-dependent mapping too can be ruled out (e.g., by
laws of quantum mechanics), and this was demonstrated in Refs. \cite{Leggett20031}
and \cite{Zeilinger2007}.

In terms of counterfactual values, the construction (\ref{eq: with c})
does introduce them implicitly, adhering thereby to counterfactual
definiteness. However, it does not require counterfactual identity,
i.e., it is not necessary that $S_{q}^{c}=S_{q}^{c'}$ whenever $q\in Q_{c}\cap Q_{c'}$
(e.g., $R_{q}^{c}$ and $R_{q}^{c'}$ are independent if so are $G_{c=1},G_{c=2},G_{c=3}$).
The assumption of counterfactual identity is equivalent to that of
context-independent mappings, this assumption is restrictive and may
be empirically violated.

It is perhaps useful here to dispel the naive but not infrequent misconception
that the random variable in ``realist'' representation must be ``single'',
so that, e.g., $G$ in (\ref{eq: U3}) is not a legitimate choice.
Any set of jointly distributed random variables can always be replaced
with a ``single'' one. In the special case of a countable set of
random variables defined on reals endowed with Borel sigma-algebra
this random variable can always be chosen, if one so wishes, as a
single variable uniformly distributed between 0 and 1 \cite{Kechris1995,DzhGluh,BranderburgerKeisler2018}.
``Probabilistic dimensionality'' (the number of components of a
random entity) is entirely a matter of one's choice. 

The obvious statement that jointly distributed random variables $X$
and $Y$ are functions of one and the same random variable $Z$ becomes
even more obvious if one realizes that $Z$ can always be chosen as
$\left(X,Y\right)$. In particular, a coupling $S$ of a system can
itself be viewed as a hidden variable, and the most ``economically
chosen'' one at that. Thus, in the classical proof of the Bell theorem
\cite{Bell1964}, for three dichotomous random variables $A,B,C$,
the hidden variable $\lambda$ could be chosen, with no loss of generality,
as the eight-valued $\lambda=\left(A,B,C\right)$. To see that this
is the most economic choice, note that any $\lambda$ such that $A=f_{A}\left(\lambda\right),B=f_{B}\left(\lambda\right),C=f_{C}\left(\lambda\right)$
can be presented as the jointly distributed quadruple $\left(A,B,C,\lambda\right)$
of which $A,B,C$ are the first three projections. Obviously, no choice
of $\lambda$ can make this quadruple simpler than eliminating $\lambda$
altogether. 

\section{Conclusion}

The language of probabilistic couplings used in CbD is a rigorous
and parsimonious way of talking about counterfactuals and hidden variables
with context-(in)dependent mapping. It is also conducive to expanding
the sphere of applicability and depth of contextuality analysis \cite{DzhKuj2014,DzhKuj2016,DzhKuj2017Fortsch,Dzh2017Nothing,DzhCerKuj2017}.
If an identically connected coupling of (\ref{eq: matrix 3}) does
not exist, other couplings do, and one can profitably study this set
of possible couplings, e.g. to compute the degree of contextuality.
Using our example (\ref{eq: matrix 3}), if the probabilities with
which $S_{1}^{1}=S_{1}^{3}$, $S_{2}^{1}=S_{2}^{2}$, and $S_{3}^{2}=S_{3}^{3}$
in a coupling (\ref{eq: matrix 3 coupling}) cannot all be 1, one
can be naturally interested in the maximal values of the sum 
\begin{equation}
\Delta=\Pr\left[S_{1}^{1}=S_{1}^{3}\right]+\Pr\left[S_{2}^{1}=S_{2}^{2}\right]+\Pr\left[S_{3}^{2}=S_{3}^{3}\right]
\end{equation}
that can be achieved among all possible couplings (\ref{eq: matrix 3 coupling}).
Then the difference $3-\Delta$ will serve as a possible measure of
contextuality of system (\ref{eq: matrix 3}). For arbitrary consistently
connected systems, $\Delta$ is the sum of $\Pr\left[S_{q}^{c}=S_{q}^{c'}\right]$
for all $\left(c,c',q\right)$ such that $q\in Q_{c}\cap Q_{c'}$,
and the measure of contextuality is $N-\Delta$, where $N$ is the
number of all such $\left(c,c',q\right)$.

As we know, the main motivation for developing CbD, was that the language
of probabilistic couplings allows one to ``smoothly'' go beyond
the class of consistently connected systems \cite{KujDzhLar2015,KujDzhProof2016,Dzh2017Nothing,DzhCerKuj2017,DzhKuj2017Fortsch,DzhKujLar,DzhKuj2018}.
An arbitrary system $\left\{ R_{q}^{c}:q\in Q_{c},c\in C\right\} $
is considered noncontextual if and only if it has a coupling $\left\{ S_{q}^{c}:q\in Q_{c},c\in C\right\} $
in which, for any $\left(c,c',q\right)$ with $q\in Q_{c}\cap Q_{c'}$,
the probability of $S_{q}^{c}=S_{q}^{c'}$ is maximal possible. Denoting
this sum of all these maximal values by $\Delta_{0}$, the measure
of contextuality mentioned above is generalized as $\Delta_{0}-\Delta$.
It is, of course, only one of many possible measures of contextuality,
other measures being described, e.g., in Refs. \cite{DzhKuj2016,DzhCerKuj2017}.
An important extension of measures of contextuality into measures
of noncontextuality, if a system is noncontextual, is addressed in
another paper in the present issue \cite{KujDzh2019}.

\subsubsection*{Acknowledgements. }

The author is grateful to the participants of the Purdue Winer Memorial
Lectures 2018 for many fruitful discussions, and to one of the anonymous
reviewers for constructive criticism.


\begin{thebibliography}{10}
\bibitem{AbramskyBrand2011}Abramsky S, Brandenburger A 2011 The sheaf-theoretic
structure of non-locality and contextuality. New J. Phys. 13: 113036-113075. 

\bibitem{AbramBarbMans2011}Abramsky S., Barbosa R. S., Mansfield
S 2017 The contextual fraction as a measure of contextuality. Phys.
Rev. Lett. 119: 050504.

\bibitem{Araujoetal2013}Araújo M, Quintino MT, Budroni C, Cunha MT,
Cabello A 2013. All noncontextuality inequalities for the n-cycle
scenario, Phys. Rev. A 88: 022118. 

\bibitem{ariasetal2015}Arias M, Canas G, Gomez ES, Barra JB, Xavier
GB, Lima G, D\textquoteright Ambrosio V, Baccari F, Sciarrino F, Cabello
A 2015 Testing noncontextuality inequalities that are building blocks
of quantum correlations. Phys. Rev. A 92: 032126.

\bibitem{Bell1964}Bell J. 1964 On the Einstein-Podolsky-Rosen paradox.
Physics, 1: 195-200.

\bibitem{Bell1966}Bell J. 1966 On the problem of hidden variables
in quantum mechanics. Rev. Mod. Phys. 38: 447-453.

\bibitem{Bohr1935}Bohr N. 1935 Can quantum-mechanical description
of physical reality be considered complete? Phys. Rev., 48: 696-702.

\bibitem{BranderburgerKeisler2018}Brandenburger A, Keisler HJ 2018
A canonical hidden-variable space. Annals of Pure and Applied Logic
169: 1295-1302.

\bibitem{BudroniBook2016}Budroni C. 2016 Temporal Quantum Correlations
and Hidden Variable Models. Springer: Heidelberg.

\bibitem{Cabello2013}Cabello A 2013 Simple explanation of the quantum
violation of a fundamental inequality. Phys. Rev. Lett., 110: 060402.

\bibitem{Cereceda2000}Cereceda J 2000 Quantum mechanical probabilities
and general probabilistic constraints for Einstein--Podolsky--Rosen--Bohm
experiments. Found. Phys. Lett. 13: 427--442.

\bibitem{crespietal2017}Crespi A, Bentivegna M, Pitsios I, Rusc D,
Poderini D, Carvacho G, D\textquoteright Ambrosio V, Cabello A, Sciarrino
F, Osellame R 2017 Single-photon quantum contextuality on a chip.
ACS Phot. 4: 2807-2812.

\bibitem{Dzh2017Nothing}Dzhafarov EN 2017 Replacing nothing with
something special: Contextuality-by-Default and dummy measurements.
In A. Khrennikov \& T. Bourama (Eds.), Quantum Foundations, Probability
and Information. Springer.

\bibitem{DzhKuj2014}Dzhafarov EN, Kujala JV 2014 Contextuality is
about identity of random variables. Phys. Scr. T163: 014009.

\bibitem{DzhKuj2016}Dzhafarov EN, Kujala JV 2016 Context-content
systems of random variables: The Contextuality-by- Default theory.
J. Math. Psych. 74: 11-33.

\bibitem{DzhKuj2017Fortsch}Dzhafarov EN, Kujala JV 2017 Probabilistic
foundations of contextuality. Fort. Phys. 65: 1-11.

\bibitem{DzhKuj2018}Dzhafarov EN, Kujala JV 2018 Contextuality analysis
of the double slit experiment (with a glimpse into three slits). Entropy
20: 278; doi:10.3390/e20040278.

\bibitem{DzhCerKuj2017}Dzhafarov EN, Cervantes VH, Kujala JV 2017
Contextuality in canonical systems of random variables. Phil. Trans.
Roy.Soc. A 375: 20160389.

\bibitem{DzhGluh}Dzhafarov EN, Gluhovsky I 2006 Notes on selective
influence, probabilistic causality, and probabilistic dimensionality.
J. Math. Psych. 50: 390--401.

\bibitem{DzhKujLar}Dzhafarov EN, Kujala JV, Larsson, J-Å 2015 Contextuality
in three types of quantum-mechanical systems. Found. Phys. 7: 762-782.

\bibitem{EPR1935}Einstein A, Podolsky B, Rosen N 1935 Can quantum-mechanical
description of physical reality be considered complete? Phys. Rev.,
47: 777-780.

\bibitem{Fine1982}Fine A 1982 Hidden variables, joint probability,
and the Bell inequalities. J. Math. Phys., Vol. 23: 1306--1310.

\bibitem{fluhmanetal2018}Flühmann C, Negnevitsky V, Marinelli M,
Home JP 2018 Sequential modular position and momentum measurements
of a trapped ion mechanical oscillator. Phys. Rev. X 8: 021001.

\bibitem{Griffiths2017}Griffiths RB 2017 What quantum measurements
measure. Phys. Rev. A 96: 032110.

\bibitem{Griffiths2019}Griffiths RB 2019 Quantum measurements and
contextuality. Phil. Trans. Roy.Soc. A, present issue.

\bibitem{Zeilinger2007}Gröblacher S, Paterek T, Kaltenbaek R, Brukner
\v{C}, \.{Z}ukowski M, Aspelmeyer M, Zeilinger A 2007 An experimental
test of non-local realism. Nature 446: 871--875.

\bibitem{Kechris1995}Kechris AS 1995 Classical Descriptive Set Theory.
New York: Springer.

\bibitem{Khrennikov2005}Khrennikov A 2005 The principle of supplementarity:
A contextual probabilistic viewpoint to complementarity, the interference
of probabilities, and the incompatibility of variables in quantum
mechanics. Found. Phys., 35: 1655 - 1693.

\bibitem{Khrennikov2009}Khrennikov A 2009 Contextual approach to
quantum formalism. Springer: Berlin.

\bibitem{Klyachko2008}Klyachko AA, Can MA, Binicio\u{g}lu S, Shumovsky
AS 208 Simple test for hidden variables in spin-1 systems. Phys. Rev.
Lett. 101: 020403.

\bibitem{KochenSpecker1967}Kochen S, Specker EP 1967 The problem
of hidden variables in quantum mechanics. J. Math. Mech. 17: 59--87.

\bibitem{KujDzhProof2016}Kujala JV, Dzhafarov EN 2016 Proof of a
conjecture on contextuality in cyclic systems with binary variables.
Found. Phys. 46: 282-299.

\bibitem{KujDzh2019}Kujala JV, Dzhafarov EN 2019 Measures of contextuality
and noncontextuality. Phil. Trans. Roy.Soc. A, present issue.

\bibitem{KujDzhLar2015}Kujala JV, Dzhafarov EN, Larsson J-Å 2015
Necessary and sufficient conditions for extended noncontextuality
in a broad class of quantum mechanical systems. Phys. Rev. Lett. 115:
150401.

\bibitem{Kurzynski2014}Kurzynski P, Cabello A, Kaszlikowski D. 2014
Fundamental monogamy relation between contextuality and nonlocality.
Phys. Rev. Lett. 112: 100401.

\bibitem{Kurzynskietal2012}Kurzynski P, Ramanathan R, Kaszlikowski
D 2012 Entropic test of quantum contextuality. Phys. Rev. Lett. 109:
020404.

\bibitem{Leggett20031}Leggett AJ 2003 Nonlocal hidden-variable theories
and quantum mechanics: An incompatibility theorem. Found. Phys. 33:
1469

\bibitem{LeggettGarg1985}Leggett AJ, Garg A 1985 Quantum mechanics
versus macroscopic realism: Is the flux there when nobody looks? Phys.
Rev. Lett. 54: 857--860.

\bibitem{LiangSpekkensWiseman2011}Liang Y-C. Spekkens RW, Wiseman
HM 2011 Specker\textquoteright s parable of the overprotective seer:
A road to contextuality, nonlocality and complementarity. Phys. Rep.
506: 1-39.

\bibitem{malinowskietal2018}Malinowski M, Zhang C, Leupold FM, Alonso
J, Home JP, Cabello A 2018 Probing the limits of correlations in an
indivisible quantum system. Phys. Rev. A 98: 050102.

\bibitem{SuppesZanotti1981}Suppes P., Zanotti M 1981 When are probabilistic
explanations possible? Synthese 48: 191--199.

\bibitem{Thorisson}Thorisson H 2000 Coupling, Stationarity, and Regeneration.
Springer, New York.

\bibitem{William}William S 2019 Counterfactuals. The Stanford Encyclopedia
of Philosophy (Spring 2019 Edition), Edward N. Zalta (ed.), URL =
https://plato.stanford.edu/archives/spr2019/entries/counterfactuals/.

\bibitem{zhanetall2017}Zhan X, Kurzy\'{n}ski P, Kaszlikowski D, Wang
K, Bian Zh, Zhang Y, Xue P 2017 Experimental detection of information
deficit in a photonic contextuality scenario. Phys. Rev. Lett. 119:
220403.
\end{thebibliography}
\end{document}